\documentclass[runningheads]{llncs}
\usepackage{graphicx}
\usepackage[table]{xcolor}
\usepackage{enumerate}
\usepackage{booktabs}
\usepackage{amssymb}
\usepackage{subcaption}
\usepackage{mwe}
\usepackage{ragged2e}
\usepackage{wrapfig}
\usepackage{booktabs}
\usepackage{multirow}
\usepackage{diagbox}
\usepackage{subcaption}
\begin{document}
\title{VTLayout: Fusion of Visual and Text Features for Document Layout Analysis}
%
\titlerunning{VTLayout: Fusion of Visual and Text Features for DLA}
%
\author{
Shoubin Li\inst{1,2} \and
Xuyan Ma\inst{1,2} \and
Shuaiqun Pan\inst{2} \and
Jun Hu\inst{2} \and 
Lin Shi\inst{2} \and
Qing Wang\inst{2}
}
%
%
\institute{University of Chinese Academy of Sciences, Beijing, China \and
The Institute of Software, Chinese Academy of Sciences, Beijing, China\\
\email{\{shoubin, hujun, shilin, wq\}@iscas.ac.cn,\\
maxuyan@bjtu.edu.cn, shuaiqun@ualberta.ca}\\
}
\maketitle              
\begin{abstract}
Documents often contain complex physical structures, which make the Document Layout Analysis (DLA) task challenging. As a pre-processing step for content extraction, DLA has the potential to capture rich information in historical or scientific documents on a large scale. Although many deep-learning-based methods from computer vision have already achieved excellent performance in detecting \emph{Figure} from documents, they are still unsatisfactory in recognizing the \emph{List}, \emph{Table}, \emph{Text} and \emph{Title} category blocks in DLA. This paper proposes a VTLayout model fusing the documents' deep visual, shallow visual, and text features to localize and identify different category blocks. The model mainly includes two stages, and the three feature extractors are built in the second stage. In the first stage, the Cascade Mask R-CNN model is applied directly to localize all category blocks of the documents. In the second stage, the deep visual, shallow visual, and text features are extracted for fusion to identify the category blocks of documents. As a result, we strengthen the classification power of different category blocks based on the existing localization technique. The experimental results show that the identification capability of the VTLayout is superior to the most advanced method of DLA based on the PubLayNet dataset, and the F1 score is as high as 0.9599.

\end{abstract}
\section{INTRODUCTION}
With the development of science and technology, more and more scientific achievements are published, and the abundant academic literature makes it difficult for scientists to extract cutting-edge innovations. Therefore, it is highly critical to extract the information needed by researchers effectively and accurately from a large amount of scientific literature. As a pre-processing step of the document understanding system, the high-performance DLA model can accurately locate and identify different category blocks in the documents. In practice, a good DLA result can improve the performance for document retrieval, text recognition, and other tasks in the field of natural language processing (NLP).

Based on the observation of the images from the PubLayNet dataset\cite{zhong2019publaynet}, the layout usually contains five categories: \emph{Figure}, \emph{List}, \emph{Table}, \emph{Text}, and \emph{Title}. According to the recent literature review, the recognition of the \emph{List} can be a challenge which the previous methods always can not perform well on \emph{List} compared with other categories \cite{zhong2019publaynet}. Besides, identifying the \emph{Title} is also one of the most difficult tasks because \emph{Title} always appears with fewer words. Therefore, to fully understand the content of scientific literature, automatically identifying the layout of document structure well has become a top priority.

The current widespread object detection and classification approaches often rely on Deep Convolutional Neural Networks (DCNNs) to obtain features. Although the \emph{Figure}, \emph{List}, and \emph{Table} in DLA are different from the objects in traditional object detection tasks, some deep-learning-based models can still perform well, such as Faster R-CNN \cite{ren2015faster}, Mask R-CNN \cite{zhong2019publaynet}, SSD \cite{liu2016ssd}, and YOLO \cite{redmon2016you}. Based on the observations of the images from the PubLayNet dataset, each category has its unique feature. Therefore, we believe a DCNNs-based approach can extract these unique features for fusion to effectively enhance the performance of DLA. Besides, some intuitive perceptual features can help classify the category blocks for the document pages with a single background color. For example, people often notice that the \emph{Title} usually presents in a bolder form than other text paragraphs. Based on these observations, we believe that the statistical pixel values of each category block can be used as a feature to classify different category blocks. Therefore, this intuitive perceptual feature is considered in our proposed model, also called the shallow visual feature in this paper. The ablation experiments prove that the extraction and recognition of shallow visual features can effectively enhance the recognition of different category blocks.

Inspired by Asim M N~\cite{asim2019two}, the model we finally propose that considers not only the visual features but also the text features of the documents. In order to improve the classification power based on the previous methods, the Faster R-CNN and Mask R-CNN models are reproduced to find out which categories are easily incorrectly classified. It can be found that the \emph{Title} can not be recognized well by both Faster R-CNN and Mask R-CNN models based on the PublayNet dataset. Therefore, after considering the text features of the \emph{Title}, we decide to apply a traditional feature extraction technique in text mining as one of the feature extraction techniques in our final proposed model.

Based on the motivations mentioned above, a novel two-stage model for DLA is proposed. The first stage is designed to locate each category block accurately. In the second stage, different category blocks are classified, including three different units related to the deep visual feature, shallow visual feature, and text features. In order to extract deep visual features from different category blocks, MobileNetV2 \cite{sandler2018mobilenetv2}, a lightweight DCNNs, is applied in our proposed model. Furthermore, the shallow visual feature is extracted based on the statistical pixel values of each category block. In addition, Term Frequency - Inverse Document Frequency (TF-IDF) \cite{ray2012domain} is also applied as a weighting method to reinforce the distinction between text format category blocks.

The contributions of our work can be summarized as follows:
\begin{itemize}
\item A document layout analysis model, VTLayout, based on the fusion of deep visual, shallow visual, and text features, is proposed to solve the low recognition rate of different document category blocks.
\item A Shallow Visual Feature Extractor is proposed to obtain intuitive perceptual features from document images.
\item Experimental results show that the VTLayout model achieves the state-of-the-art performance on the PubLayNet dataset.
\end{itemize}

The paper is structured as follows: In this section, the motivations of DLA and contributions of this paper are introduced. The second section summarizes the latest literature review in DLA. Then, the third section briefly describes the structure of the VTLayout model with different feature extractors in detail, and the PubLayNet dataset is introduced in section 4. In the next section, the experiment settings are listed in detail and we present the experimental results of the VTLayout model on the PubLayNet dataset with some further analysis in section 6. The last section concludes the VTLayout model and future research in DLA. 

\section{RELATED WORK}
Scientists have already proposed some methods for DLA. For example, a method for page layout analysis has been proposed based on the bottom-up, nearest-neighbor clustering of page components \cite{o1993document}. Meanwhile, this method generated precise measures of skew, within-line, and between-line spacings. Besides, text lines and blocks were also located. Another traditional page segmentation technique has been proposed based on the recursive X-Y \cite{ha1995recursive}. Moreover, the black pixels were used for connected components instead of using image pixels. In 2007, Namboodiri and Jain \cite{namboodiri2007document} proposed a workflow of the document layout and structure analysis system that includes the pre-processing, layout, and structure analysis, segmented document, and evaluation steps. 

In recent years, Zhong et al. \cite{zhong2019publaynet} published a huge dataset named the PubLayNet. It was created for DLA by automatically matching the XML representation, and the content includes more than one million PDF articles publicly available on PubMed Central. Based on the PubLayNet dataset, three more experiments were also made to investigate. Firstly, it has been demonstrated that Faster R-CNN and Mask R-CNN models can perform well on this dataset, although there was much room for improvement. Next, the Faster R-CNN and Mask R-CNN models were pre-trained on PubLayNet by researchers and fine-tuned successfully to tackle the ICDAR 2013 Table Recognition Competition. Thirdly, experimental results have demonstrated that the PublayNet dataset can be used for transfer learning in distant domains.

Furthermore, the researchers proposed a series of new methods based on the DCNNs \cite{binmakhashen2019document} \cite{augusto2017fast} \cite{soto2019visual}. For document image classification, Kang et al. \cite{kang2014convolutional} applied a DCNNs architecture on the Tobacco 3482 benchmark dataset to learn from the raw image pixels. The experimental result has demonstrated that the proposed method can surpass the performance of the simple structure-based approach. In 2018, Kavasidis et al. \cite{kavasidis2018saliency} proposed a method that combined the DCNNs, graphical models, and saliency concepts to solve the table and chart detection task. Siddiqui et al. \cite{siddiqui2019rethinking} solved the table structure recognition task from the domain of semantic segmentation. In addition, a method of prediction tiling based on the consistency assumption was proposed for the tabular structure, which achieved excellent performance on the ICDAR-13 image-based table structure recognition dataset. Sun et al. \cite{sun2019faster} proposed a table detection method based on the Faster R-CNN architecture combined with the corner location method.

In addition, some researchers believe that combining the semantic text information of the documents is a benefit for DLA. In 2017, Yang et al.~\cite{yang2017learning} proposed an end-to-end, multimodal, and fully convolutional network for document semantic structure. Meanwhile, the unified model classified pixels not only by their appearance as in the traditional page segmentation task but also by the content of the underlying text. In 2019, a novel two-stream approach was proposed based on the feature-ranking algorithm for document image classification \cite{asim2019two}. Meanwhile, an average ensembling method was applied to concatenate the textual and visual stream in the proposed approach. Jain and Wigington \cite{jain2019multimodal} proposed another method for DLA based on the multimodal feature fusion combining a feature representation of the visual and text modalities.

\section{METHODOLOGY}

\begin{figure}[h]
    \centering
    \includegraphics[width=\textwidth]{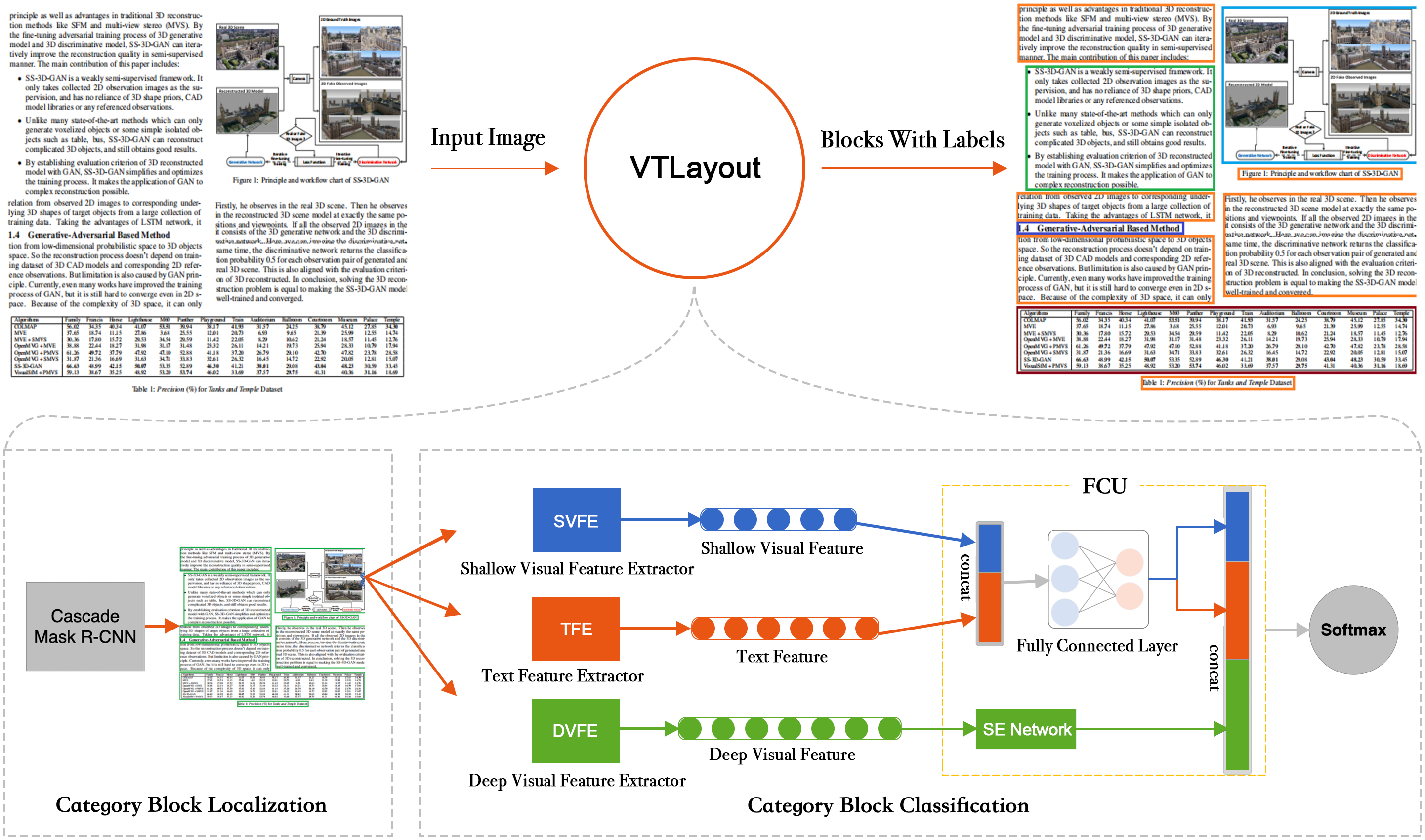}
    \caption{The structure of the VTLayout consists of two stages, Category Block Localization and Category Block Classification. The Category Block Localization stage localizes the different categories from scientific documents using the Cascade Mask R-CNN model. The DVFE, SVFE, and TFE have been built to extract different features in the Category Block Classification stage. The DVFE is built with the MobileNetV2 model to extract the deep visual feature from all the category blocks. The SVFE extracts the shallow visual feature based on the statistical pixels of different category blocks. The TFE is implemented with the TF-IDF feature extraction technique to extract the text features from the category blocks.}
    \label{fig2}
\end{figure}

This section introduces the workflow of VTLayout in detail. VTLayout consists of two stages for DLA. A proven efficient object detection model has been applied directly to localize different category blocks in the first stage. Then, a novel classification approach has been proposed in the second stage based on the fusion of the deep visual, shallow visual, and text features. Fig~\ref{fig2} exhibits the two stages of the VTLayout model.

\subsection{Category Block Localization}

In the first stage, the document images are sent to the Cascade Mask R-CNN model \cite{cai2018cascade}, where all the different category blocks are localized. The Cascade Mask R-CNN model extends the Cascade R-CNN by adding a mask head to the cascade. In object detection, the intersection over union (IoU) \cite{cai2018cascade} threshold is required to define positives and negatives, and it is expected that some existing methods can not perform well when the IoU threshold increases. Therefore, the Cascade R-CNN is proposed for solving this problem with a multi-stage object detection architecture trained with increasing IoU thresholds.

\subsection{Category Block Classification}

Although it is demonstrated that Cascade Mask R-CNN is shown to surpass all the single-model object detectors on the challenging COCO dataset \cite{lin2014microsoft}, due to the similarity of \emph{List}, \emph{Text}, \emph{Title}, it can locate each category block rather well but can not classify them correctly in the PubLayNet dataset. Therefore, in the second stage, a novel approach is proposed to improve the classification power of the category blocks based on the fusion of the deep visual, shallow visual, and text features. The Deep Visual Feature Extractor (DVFE) is built with the MobileNetV2 model to extract the deep visual feature from all the category blocks. The shallow visual feature is also extracted by the Shallow Visual Feature Extractor (SVFE) based on the statistical pixels of different category blocks. To extract the text features from the category blocks, the TF-IDF feature extraction technique is applied as the primary technology in the Text Feature Extractor (TFE). Then, a Squeeze-and-Excitation (SE) network \cite{hu2018squeeze} is applied with the MobileNetV2 model to weigh each feature map extracted by the MobileNetV2 model. Meanwhile, the extracted shallow visual feature and text feature vectors are concatenated and sent to a fully connected layer for further classification. The following subsections, the DVFE, SVFE, TFE, and Feature Concatenation Unit (FCU), are introduced with more details.

\subsubsection{Deep Visual Feature Extractor}
\ 
\newline
MobileNetV2 model is selected as the backbone architecture, which has been proved to achieve excellent results in multiple tasks and benchmarks. Compared with other proposed architectures, MobileNetV2 is a lightweight DCNNs architecture, which delivers high accuracy results with small numbers of parameters and mathematical operations. The basic structure of the MobileNetV2 is a bottleneck depth-separable convolution with residuals and contains the initial fully convolution layer \cite{sandler2018mobilenetv2}. Based on the MobileNetV1 model \cite{howard2017mobilenets}, researchers found that removing non-linearities in the narrow layers is essential to maintain the representational power and the experimental results proved its feasibility.

\subsubsection{Shallow Visual Feature Extractor}
\begin{figure}
    \centering
    \includegraphics[width=0.9\textwidth]{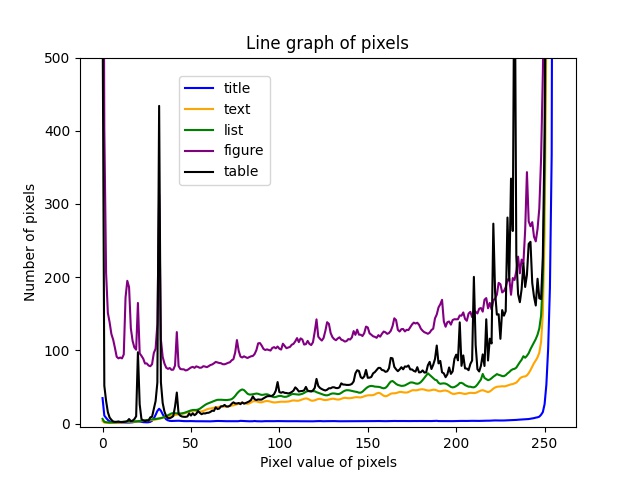}
    \caption{Statistical pixels of different category blocks.}
    \label{figtitle}
\end{figure}

\ 
\newline
Based on the experimental results from the Category Block Localization stage, \emph{List}, \emph{Text} and \emph{Title} can be misclassified by the Cascade Mask R-CNN. For most images, pixel values range from 0 (black) to 255 (white). As Fig. \ref{figtitle} shows, the statistical pixels of \emph{Figure}, \emph{List}, \emph{Table}, \emph{Text} and \emph{Title} are plotted as line graphs for comparison based on all the category blocks from the PubLayNet training dataset. The horizontal axis represents the pixel values range from 0 to 255, and the vertical axis represents the number of pixels for each pixel value. Based on the line graph of pixels, it is easy to find that the pixel values between 0 to 255 have distinct characteristics to classify different category blocks. Therefore, the SVFE is proposed as one of the units, and all the feature vectors are in length 256.

\subsubsection{Text Feature Extractor}
\ 
\newline
Text Feature Extractor (TFE) is built for extracting the text features of all the different category blocks. Since every word needs to be extracted for text features, all the category blocks are applied with the PaddleOCR \cite{shi2016end}. In particular, as \emph{Title} blocks are always in small formats and even difficult to be identified by human naked eyes, we particularly enlarge each category block eightfold. Then, the TF-IDF feature extraction technique is applied with the output of PaddleOCR to determine the importance of each word in the textual content. Finally, a vector is built for each category block which represents the text feature.

\subsubsection{Feature Concatenation Unit}
\ 
\newline
The DVFE can extract important feature maps that cannot be directly observed, while SVFE can extract features that may be missed by the network in the process of convolution. Furthermore, the features extracted by TFE can enhance the discriminant capability of text format category blocks. Considering the three kinds of extracted features, increasing the weights of the more valuable features by the whole classifier automatically becomes the next research key point. Researchers have found that enhancing the quality of spatial encodings throughout its feature hierarchy is a proper way to strengthen the representational power of CNN. In this work, the SE block has been proposed to recalibrate the channel-wise feature adaptively \cite{hu2018squeeze}. Since the SE block has been proved that it can improve the performance of CNNs at a small additional computational cost, we apply the SE block with the MobileNetV2 model.

Besides the deep visual feature, the shallow visual feature and the text features are merged into a single larger vector. Then, the concatenated vector is put into a four-layer, fully connected deep neural networks (DNNs) with the number of neurons [512, 256, 128, 64], and the output is concatenated with the output of the SE block.

\section{DATASET}
This section introduces the dataset used in our experiments in detail. The PubLayNet dataset is the largest dataset ever for DLA task. It contains more than 360,000 document images with five annotated document layout categories. Table \ref{Tab03} shows the statistics of the whole PubLayNet dataset, which includes the training and validation dataset. As Table \ref{Tab03} shows, the \emph{Text} category contains an enormous amount of data compared to other categories, which is expected because the amount of \emph{Text} is often much more extensive than other categories in scientific literature.

Meanwhile, it is easy to find that the amount of data in \emph{Title} is ranked as the second, and the number of \emph{List} is the smallest in the entire dataset. For experiments of reproducing the baselines and the VTLayout models, all the images from the PubLaynet dataset are used. As the test dataset is not published so far, we use all the images from the training dataset for training, and the whole validation dataset is used for inference. Therefore, there are 335,703 images for training and 11,245 images for inference.

\begin{table}
\renewcommand\arraystretch{1.2}
\centering
\caption{Statistic of the PubLayNet dataset.}
\label{Tab03}
\begin{tabular}{|c|c|c|}
\hline
Categories of the dataset & Training dataset & Validation dataset\\
\hline
Text & 2,376,702 & 88,625 \\
Title & 633,359 & 18,801 \\
List & 81,850 & 4,239 \\
Table & 103,057 & 4,769\\
Figure & 116,692 & 4,327  \\ \hline
Total & 3,311,660 & 120,761 \\
\hline
\end{tabular}
\end{table}

\section{EXPERIMENTAL SETTINGS}
This section describes all the experimental settings of this paper. In the beginning, we compare the performance of our proposed VTLayout model with the Faster R-CNN, Mask R-CNN baselines published in the paper of PubLayNet dataset \cite{zhong2019publaynet}. Besides, we also compare the Cascade Mask R-CNN experimental results with the VTLayout model to observe the effectiveness of fusing the visual and text features for DLA. As we want to compare the classification power of the VTLayout model with the baselines, we reproduce the baselines based on their experimental settings and analyze their classification results without localization results. Therefore, precision, recall, and F1 score \cite{goutte2005probabilistic} are applied as the primary evaluation metrics to evaluate the models in this experiment instead of the MAP @ IOU [0.50:0.95] evaluation metric applied in the paper of PubLayNet dataset.

In our VTLayour model, the Cascade Mask R-CNN model is implemented by Pytorch framework \cite{paszke2019pytorch}. Then, the resNeXt-101-64x4d model is selected as the backbone network, initialized with the model pre-trained on the ImageNet dataset. The model is trained for 30 epochs with a batch size of 8, one sample per GPU. Moreover, SGD optimizer \cite{keskar2017improving} is used with the initial learning rate of 0.02, the momentum of 0.9, and the weight-decay of 0.0001. In DVFE, the mobileNetV2 model is applied with the TensorFlow framework \cite{abadi2016tensorflow} and all the category blocks are resized to 128 x 128 by padding. The mobileNetV2 is pre-trained by the ImageNet dataset without the fully connected layer at the top of the network. Besides, the three input channels are set as (128, 128, 3), and the global average pooling is applied to the output of the last convolutional block. In the SVFE, all the color images are firstly converted to grayscale images. In the first step of TFE, the chinese\_ocr\_db\_crnn\_mobile \cite{zhou2017east} from PaddleOCR is applied to identify the words from the documents. Then, the TF-IDF feature extraction technique is applied with the Sklearn library \cite{pedregosa2011scikit}. Finally, the Adam optimizer \cite{keskar2017improving} is used with the initial learning rate of 0.001, and the cross-entropy is used as the loss function.

Besides comparing the VTLayout model with the baselines, a series of stability checking experiments are built based on a small-sized dataset. In order to test our proposed VTLayout model on a small dataset, we only train the Category Block Classification stage with 25,000 randomly selected images for \emph{Text} and \emph{Title}, 10,000 randomly selected images for \emph{Figure}, \emph{List} and \emph{Table} from the output of the Cascade Mask R-CNN instead of all of the training dataset. Based on the statistics of the PubLayNet training dataset, there are around seven \emph{Text} blocks, and two \emph{Title} blocks can be extracted from one page. As around four pages can contain one \emph{List}, three pages can contain one \emph{Table} and one \emph{Figure}, 30,000 images from the PubLayNet training dataset are randomly selected to train the Faster R-CNN and Mask R-CNN to see the classification capability of each model. The results of the comparative experiment are presented in the Result section. 

Meanwhile, a five-fold cross-validation experiment is also applied to test the stability of our VTLayout model on the same small-sized dataset. In the experiments, the dataset is randomly divided into five parts on average and takes out four of them as the new training dataset and the remaining one as the test dataset each time. Besides, each fold of data is required to be the test dataset once. Finally, in the ablation experiments, a series of experiments are implemented to see whether all the three feature extractors can affect the master model and what kind of feature can contribute most to the VTLayout model. In particular, if the DVFE is not implemented as one of the units in the VTLayout model, the SE network will also not be applied. If the SVFE or TFE are not applied in the VTLayout model, their feature vectors will not be concatenated and put into the fully connected layer. F1 score is selected as the only evaluation metric for the stability experiments and ablation experiments.

\section{RESULTS AND ANALYSIS}
Table \ref{Tab04} shows that our proposed VTLayout model achieves a state-of-the-art performance than the baseline models on the PubLayNet dataset with the F1 score of 0.9599. The excellent F1 score means that our model has both low false positives and low false negatives, and the low precision and recall values also prove the effectiveness of the VTLayout model.

\begin{table}
\renewcommand\arraystretch{1.2}
\centering
\caption{VTLayout performance compared with the baselines.}
\label{Tab04}
\begin{tabular}{|p{4cm}<{\centering}|p{2cm}<{\centering}|p{2cm}<{\centering}|p{2cm}<{\centering}|}
\hline
Model & Precision & Recall & F1 Score \\
\hline
Faster R-CNN  & 0.9319 & 0.9130 & 0.9224 \\
Mask R-CNN  & 0.9379 & 0.9410 & 0.9385\\
Cascade Mask R-CNN & 0.9515 & 0.9506 & 0.9510\\
VTLayout (Ours) & \textbf{0.9584} & \textbf{0.9618} & \textbf{0.9599}\\
\hline
\end{tabular}
\end{table}

Moreover, we compare the F1 scores of the Faster R-CNN, Mask R-CNN, Cascade Mask R-CNN, and our proposed VTLayout model based on each category block. As Table \ref{Tab05} shows, it can be found that our proposed VTLayout model achieve the state-of-the-art performance on identifying the \emph{Table}, \emph{Text} and \emph{List}. Compared with the Cascade Mask R-CNN model, our VTLayout model can perform better in most categories, although there is a small drop in recognition of \emph{Figure} with the F1 score of 0.9824 only. In particular, the VTLayout model successfully made up for the inaccuracy of Cascade Mask R-CNN in \emph{Title}'s recognition which increases the F1 score from 0.9166 to 0.9411, although Faster R-CNN model can work better on recognition of the \emph{Title} with the F1 score of 0.9425.  Meanwhile, we can find that the identification capability of the \emph{List} is the worst among the five categories in baselines, but our proposed model greatly improves the recognition power of the \emph{List} with the F1 score of 0.9177.

\begin{table}
\renewcommand\arraystretch{1.2}
\centering
\caption{F1 score comparison on different categories.}
\label{Tab05}
\begin{tabular}{|c|p{2cm}<{\centering}|p{2cm}<{\centering}|p{2.5cm}<{\centering}|p{2cm}<{\centering}|}
\hline
\diagbox[height=4em]{Categories}{Model} & Faster R-CNN & Mask R-CNN & Cascade Mask R-CNN & VTLayout (Ours) \\
\hline
Text & 0.9475  & 0.9475  & 0.9688 & \textbf{0.9751}  \\
Title & \textbf{0.9425} & 0.9406  & 0.9166 & 0.9411 \\
List  & 0.8150  & 0.8874 & 0.9055 & \textbf{0.9177} \\
Figure  & 0.9663 & 0.9617 & \textbf{0.9846} & 0.9824\\
Table  & 0.9376 & 0.9553 & 0.9794 & \textbf{0.9833} \\
\hline
\end{tabular}
\end{table}

Based on the paper of PubLayNet dataset, the MAP @ IOU [0.50:0.95] values of the \emph{Title} were the worst compared with other categories by Faster R-CNN and Mask R-CNN models. However, according to our reproduction of the two models, the experimental results from Table \ref{Tab05} show that \emph{Title} can be recognized better than \emph{List}. Thus, it proves that \emph{Title} can be well recognized but localizing the \emph{Title} can be challenging because it can be recognized easily as part of the \emph{Text}.

\subsubsection{Correction of the wrong cases}
\begin{figure}
    \centering
    \includegraphics[height=6cm,width=5.8cm]{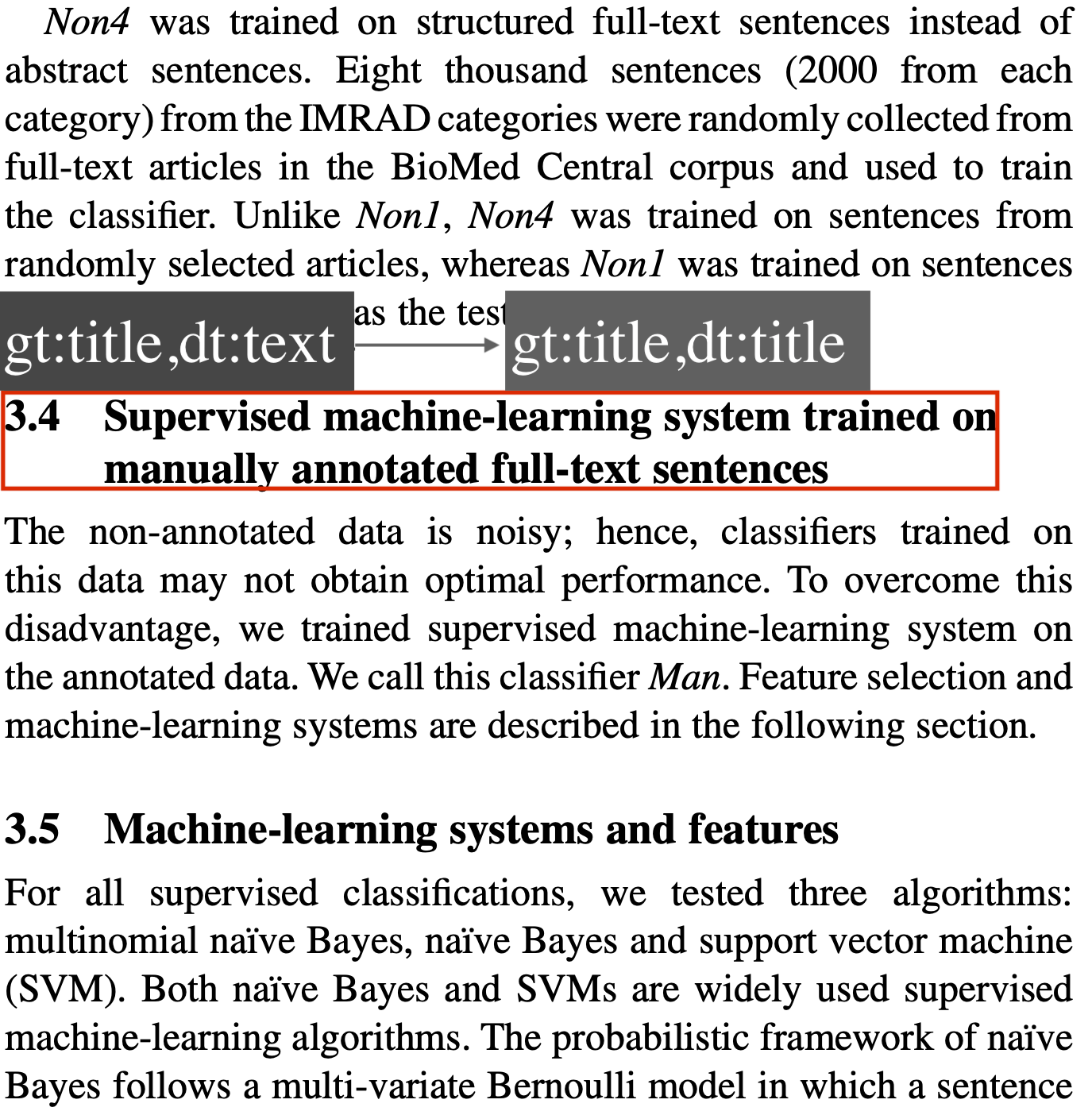}
    \includegraphics[height=6cm,width=6.2cm]{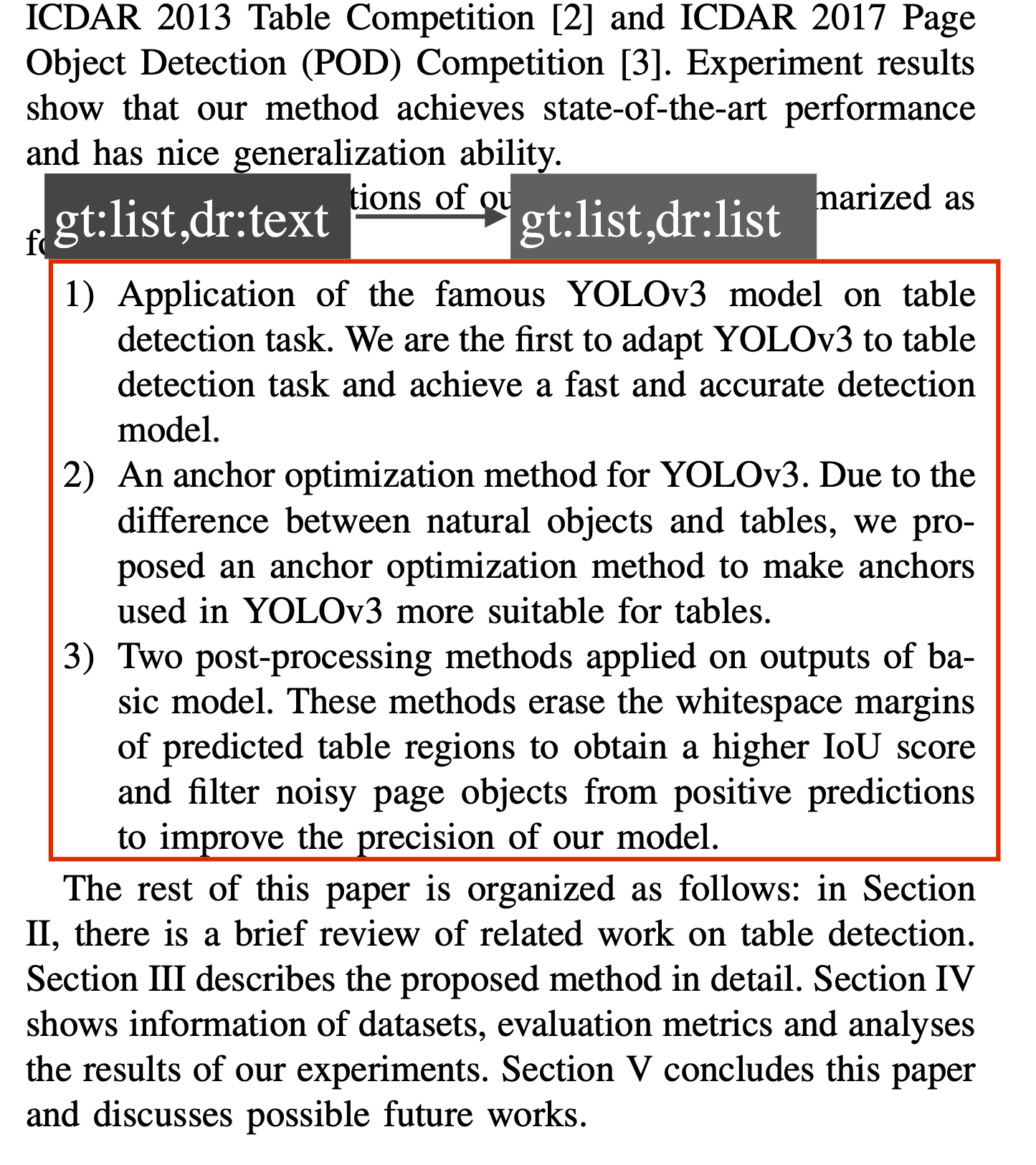}
    \caption{Corrections of the wrong cases.}
    \label{wrongcase}
\end{figure}

\
\newline
As shown in Fig~\ref{wrongcase}, two images have been shown as examples of the corrections based on our proposed VTLayout model. These two images are wrong predictions by the Cascade Mask R-CNN. On the left-hand side, the \emph{Title} is recognized wrongly by the Cascade Mask R-CNN, and the VTLayout model predicts the \emph{Title} correctly. On the right-hand side, Cascade Mask R-CNN predicts the \emph{List} as \emph{Text}, but our VTLayout model corrects this error successfully.

\subsubsection{Stability checking on small-sized dataset}
\
\newline
The experiment of checking the VTLayout model on the small-sized dataset shows that the overall F1 score decreases slightly from 0.9599 to 0.9546 on the whole validation dataset. For the stability checking experiment on Faster R-CNN, the F1 score decreases from 0.9224 to 0.9168 with a drop around 0.0056. Meanwhile, the F1 score of Mask R-CNN also decreases from 0.9385 to 0.9361, with a drop of around 0.0024. These experimental results demonstrate that our proposed VTLayout model has excellent stability on the small-sized dataset as well as the Faster R-CNN and Mask R-CNN model. 

Table \ref{tab:table3} shows the experimental results of the five-fold cross-validation experiment, which demonstrates that the F1 score of all kinds is relatively stable. Overall, all the experimental results prove that the stability of the VTLayout model is high, and the contingency is low.

\begin{table}[!ht]
\caption{Results of the five-fold cross-validation experiment of F1 scores on different categories of the VTLayout model.}
\begin{tabular*}{\hsize}{@{}@{\extracolsep{\fill}}lllllll@{}}
\toprule
& Text  & List   & Title  & Figure & Table  & Average \\ \hline
Fold1 & 0.9450 & 0.9648 & 0.9295 & 0.9560 & 0.9472 & 0.9495   \\ 
Fold2  & 0.9428 & 0.9746 & 0.9056 & 0.9878 & 0.9735 &0.9569  \\ 
Fold3 & 0.9402 & 0.9760 & 0.9347 & 0.9627 & 0.9512 & 0.9532 \\ 
Fold4 & 0.9562 & 0.9810 & 0.9537 & 0.9848 & 0.9626 & 0.9678 \\ 
Fold5 & 0.9376 & 0.9687 & 0.9195 & 0.9591 & 0.9425 & 0.9458  \\ \hline
Average & 0.9444 & 0.9286 & 0.9730 & 0.9701 & 0.9554 & 0.9546 \\ 
\bottomrule
\end{tabular*}
\label{tab:table3}
\end{table}

\subsubsection{Ablation experiments}
\
\newline
Table \ref{ablation} shows the F1 scores of the VTLayout model and all the ablation experiments, in which DVFE denotes the Deep Visual Feature Extractor, SVFE represents the Shallow Visual Feature Extractor, and TFE represents the Text Feature Extractor. Meanwhile, VTLayout$_{DVFE+SVFE+TFE}$ represents our proposed VTLayout model, and VTLayout$_{DVFE+SVFE}$ denotes the VTLayout model without the Text Feature Extractor. The experimental results demonstrate that all the components impact the master model because the loss of any component results in performance degradation. In particular, the VTLayout$_{DVFE}$ experiment shows that although DVFE can work well on the recognition of the \emph{Text} with an F1 score of 0.9789, the loss of SVFE and TFE can result in a decrease of F1 score on identifying the \emph{List} to 0.7881. Besides, we also can conclude that the DVFE makes the most significant contribution to our model. While TFE has the worst recognition rate for category blocks, it is still necessary because it can make the overall performance of the model better.

Overall, based on the experimental results and analysis above, we demonstrate that our proposed VTLayout model is superior to the current most advanced methods of DLA. The reasons can be summarized as follows. Firstly, our VTLayout model fuses three different kinds of features, deep visual, shallow visual, and text features from the PublayNet dataset, which can boost the performance of identifying different category blocks. In particular, the experimental results of the ablation experiments have demonstrated that all three features are beneficial and necessary for DLA. Secondly, with the great success of deep learning in object detection, researchers have neglected the importance of traditional shallow visual features. Our proposed extractor for the shallow visual feature is one of the most important contributions in this paper. Thirdly, the experimental results prove that the VTLayout model has excellent stability on randomly selected small datasets. 

In addition, our proposed method still has some deficiencies, which the recognition of \emph{Title} and \emph{List} need to be further improved. Based on these findings and recent literature review, the recognition of \emph{List} still remains challenging compared with other different category blocks, and it can be predicted wrongly as \emph{Text} category easily. Therefore, improving the identification rate of the \emph{List} has become our next focus, and we believe that the bullet points in the \emph{List} can be the key to solving this problem. One of the main reasons for inaccurate recognition of \emph{Title} is that many of the \emph{Titles} in the text have only one or two words, so the \emph{Titles} are often appear in small sizes and even the human cannot accurately identify them with naked eyes. In the experiments, although we enlarge the size of \emph{Title} blocks before applying with the PaddleOCR, the blurring caused by the amplification still makes the recognition rate of \emph{Title} less than ideal.

\begin{table*}[]
\renewcommand\arraystretch{1.3}
\centering
\caption{The ablation experimental results of F1 scores of the VTLayout model.}
\label{ablation}
\begin{tabular}{|p{0.5cm}<{\centering}|p{4.3cm}<{\centering}|p{1cm}<{\centering}|p{1cm}<{\centering}|p{1cm}<{\centering}|p{1cm}<{\centering}|p{1cm}<{\centering}|p{1.3cm}<{\centering}|}
\hline
No. & Model & Text   & Title   & List  & Figure & Table & Average \\
\hline
1 & VTLayout$_{DVFE+SVFE+TFE}$ &  0.9751 & 0.9411 & 0.9177 & 0.9824 & 0.9833 & 0.9599\\
2 & VTLayout$_{DVFE+SVFE}$ & 0.9638 & 0.9296 & 0.8725 & 0.9834 & 0.9678 & 0.9440\\
3 & VTLayout$_{DVFE+TFE}$ & 0.9230 & 0.8633 & 0.7635 & 0.9625 & 0.9656 & 0.8956 \\
4 & VTLayout$_{SVFE+TFE}$ & 0.8412 & 0.7699 & 0.6944 & 0.9308 & 0.9261 & 0.8455 \\
5 & VTLayout$_{DVFE}$ & 0.9789 & 0.9273 & 0.7881 & 0.9801 & 0.9576 & 0.9272\\
6 & VTLayout$_{TFE}$ & 0.3657 & 0.3635 & 0.1030 & 0.1990 & 0.1505 & 0.3198 \\
7 & VTLayout$_{SVFE}$ & 0.8913 & 0.8391 & 0.3753 & 0.9275 & 0.9273 & 0.8209 \\
\hline
\end{tabular}
\end{table*}

\section{CONCLUSION}
A VTLayout model is proposed for DLA task based on the fusion of deep visual, shallow visual, and text features. The experimental results show that the proposed VTLayout model is superior to the most advanced classification methods in the PublayNet dataset, and the F1 score is 0.9599. Meanwhile, we find that the intuitive perceptual feature is beneficial to the DLA. As we mentioned in the introduction, the accuracy of the DLA can determine the performance of many NLP-related tasks. An accurate and efficient DLA model can accurately locate a category block and extract it from many complex datasets for future work. As far as I know, in the field of applied chemistry, some researchers prefer to pay more attention to the \emph{Tables} only from the literature. Our work could significantly shorten the time it takes researchers to find \emph{Tables} in thousands of academic papers. In addition, DLA can be beneficial to the evaluations of grant applications. In recent years, government and top research institution funding agencies gradually started to apply AI techniques to assist manual evaluating of grant applications. Our proposed VTLayout model has been applied to extract the \emph{Text} and \emph{Tables} from thousands of grant applications for further evaluation.

Although the VTLayout model achieves the state-of-the-art performance in identifying different categories in DLA, there is still much room for improvement in both localization and classification of different category blocks. Firstly, compared with the traditional DCNN-based object detection models, Transformer-based backbones start to show up and achieve remarkable results in a series of traditional public datasets for object detection. Therefore, applying the Transformer-based method to DLA will be our next research direction. Secondly, assign features that have more influence on classifier performance with more weights can be another key consideration in the future. We believe that the model's performance can be improved by self-adjusting the weights of deep visual, shallow visual, and text features. Next, the accuracy of OCR is self-evident for the VTLayout model. In recent years, the research of scene text recognition has made significant progress. Complex background conditions, text color, font size, and irregular text representation are no longer obstacles in recognizing the scene text. Therefore, we believe that the achievements of scene text recognition can meet our requirements for OCR's accuracy. Fourthly, we will continue to explore further the research of multimodal fusion in DLA to optimize the extraction of deep visual, shallow visual and text features of documents. In addition, we also seek to come up with an end-to-end, lighter DLA model.

\clearpage
\bibliographystyle{splncs04}
\bibliography{main}

\end{document}